\documentclass[prb,amsmath,amssymb,twocolumn,showpacs]{revtex4}

\usepackage{graphicx}
\usepackage[usenames]{color}
\usepackage[english]{babel}
\usepackage{ulem}

\usepackage[latin1]{inputenc}

\usepackage[T1]{fontenc}

\begin{document}

\title{$^{63,65}$Cu NMR and NQR evidence for an unusual spin dynamics in PrCu$_2$ below 100~K}
\author{A.~Sacchetti$^1$, M.~Weller$^1$,
J.~L.~Gavilano$^2$, R.~Mudliar$^{1,3}$, B.~Pedrini$^{1,4}$,
K.~Magishi$^{1,5}$, H.~R.~Ott$^1$, R.~Monnier$^1$, B.~Delley$^6$,
Y.~\={O}nuki$^7$}

\affiliation{ $^1$ Laboratorium f\"{u}r Festk\"{o}rperphysik,
ETH-H\"{o}nggerberg, CH-8093~Z\"{u}rich\\
$^2$ Laboratory for Neutron Scattering, Paul Scherrer Institut,
CH-5232 Villigen PSI, Switzerland\\$^3$Indian Institute of
Technology, Bombay Powai, Mumbai - 400076, INDIA\\ $^4$Department
of Molecular Biology The Scripps Research Institute 10550 North
Torrey Pines Road, MB-44 La Jolla, CA 92037 USA\\
$^5$Faculty of Integrated Arts and Sciences, The University of
Tokushima, Tokushima 770-8502, Japan\\$^6$ Paul Scherrer Institut,
CH-5232 Villigen PSI, Switzerland\\$^7$ Faculty of Science, Osaka
University, Machikaneyama, Toyonaka, Osaka 560, Japan}

\date{\today}

\begin{abstract}
We report the results of a $^{63,65}$Cu NMR/NQR study probing the
intermetallic compound PrCu$_2$. The previously claimed onset of
magnetic order at 65~K, indicated in a $\mu$SR study, is not
confirmed. Based on our data we discuss different possible reasons
for this apparent discrepancy, including a non negligible
influence of the implanted muons on their environment. Competing
dipolar and quadrupolar interactions lead to unusual features of
the magnetic-ion/conduction-electron system, different from those
of common intermetallics exhibiting structural or magnetic
instabilities.
\end{abstract}



\maketitle

\section{Introduction}
The Van Vleck paramagnet PrCu$_2$ exhibits peculiar electronic and
magnetic properties, whose complete understanding is still to be
achieved. The $4f$ electron ground-state multiplet of the
Pr$^{3+}$ ions is split into nine singlets by the crystal electric
field. Nevertheless, an induced Jahn-Teller (JT) transition is
observed at $T_\textrm{JT}=7.6$~K,\cite{JT,Ott} where the crystal
structure changes from orthorhombic to slightly
monoclinic.\cite{StructJT,AFmK1,Jorge} An incommensurate
antiferromagnetic (AF) order below $T_N=54$~mK among the Pr
nuclear magnetic moments was claimed on the basis of magnetic
susceptibility, specific heat, and neutron scattering
measurements.\cite{AFmK1,AFmK2} The respective moments align along
the $a$ direction of the orthorhombic unit cell with a propagation
vector $\roarrow q=(0.24,0,0.68)$ in the reciprocal lattice.
Magnetization and susceptibility measurements indicate a large
polarization of the $4f$ Pr magnetic-moments in the presence of an
external field.\cite{chi,metmag} At low temperatures the easy and
hard magnetization axes are along the $a$ and $c$ direction,
respectively. Quite surprisingly the two magnetization axes are
exchanged for fields exceeding 10~T, leading to a metamagnetic
transition.\cite{chi,metmag} The $4f$ electron Pr-quadrupole
moments play an important role in the physics of PrCu$_2$. In
particular the JT transition is ascribed to a phonon-mediated
interaction between the quadrupole moments,\cite{StructJT} whereas
the metamagnetic transition has been ascribed to a change in the
orientation of the spatial charge distribution.\cite{CEF}

Based on the results of a muon spin resonance ($\mu$SR) study, AF
ordering of the localized Pr $4f$ electron moments below 65~K has
recently been suggested.\cite{muons} Since such a magnetic
transition in this temperature region is not reflected in the
results of either neutron scattering or susceptibility
experiments,\cite{AFmK1,chi,StructJT} it may be assumed that the
inferred magnetic order does not involve long-range order and
static internal fields. Considering the electronic configuration
of the Pr$^{3+}$ ions, the above result is really unexpected and
calls for additional checks. Since the $\mu$SR results may be due
to dynamically-induced non vanishing internal fields, it is
important to use a microscopic technique for investigating the
phenomenon further. In this paper we report a systematic study of
PrCu$_2$ by means of Cu-based nuclear magnetic resonance (NMR) and
nuclear quadrupole resonance (NQR) experiments between 4.2~K and
room temperature.

\section{Experimental}\label{expt}
Nuclear resonance measurements were made using standard spin-echo
techniques employing a phase-coherent pulsed spectrometer.
$^{63,65}$Cu NMR and NQR spectra were collected by measuring the
integrated spin-echo signal as a function of the exciting
radio-frequency (RF) in fixed external magnetic fields. The
spin-lattice relaxation rate (SLRR) $T_1^{-1}$ was measured by
destroying the $z$-component of the equilibrium nuclear
magnetization by means of a short RF pulse and monitoring its
recovery as a function of a variable delay between this
preparation pulse and a spin-echo sequence. The spin-spin
relaxation rate (SSRR) $T_2^{-1}$ was obtained from the $\tau$
dependence of the signal after a $\pi/2-\tau-\pi$ sequence.

Single crystalline PrCu$_2$ samples were grown at Osaka University
by means of the procedure described in Ref.~\onlinecite{sample}.
Since the system is metallic, the NMR signal from the single
crystal is very small because the penetration depth of the RF
pulses is of the order of 1~$\mu$m only. We observed an unexpected
splitting and broadening of the NMR lines, which we traced back to
a non-perfect orientation of the applied magnetic field with
respect to one of the crystalline axes of the sample. From a
detailed analysis we concluded that meaningful NMR measurements on
a single crystal of PrCu$_2$ require a precision in sample
orientation that is much better than $\pm 1^{\circ}$, which could
not be achieved with our experimental set-up. For this reason we
powdered the single-crystalline sample which not only served to
avoid the orientation problem but also helped to enhance the NMR
signal.

\section{NMR measurements}\label{NMRmeas}
\begin{figure}[!tb]
\center
\includegraphics[width=8cm]{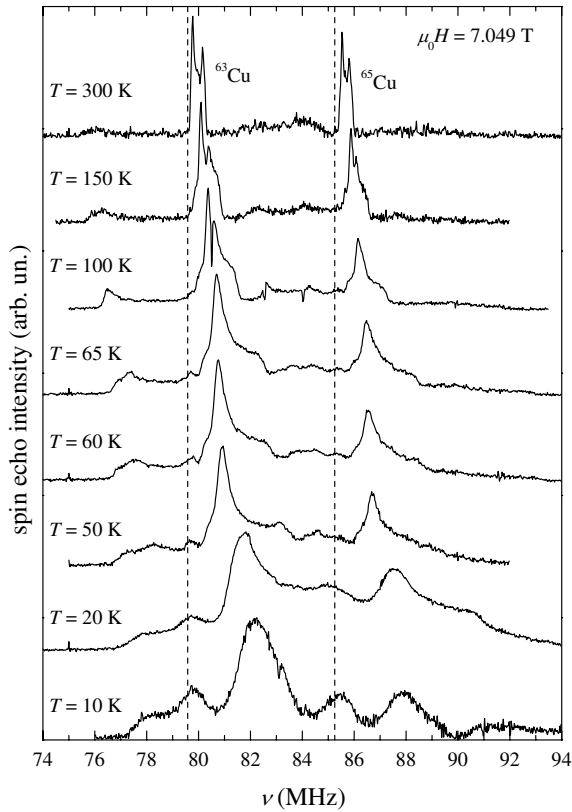}
\caption{$^{63,65}$Cu-NMR powder spectra of PrCu$_2$ at selected
temperatures and in a field of 7.049~T. The vertical dashed lines
mark the reference NMR frequencies for the two
Cu-isotopes.\cite{Bruker}} \label{NMR}
\end{figure}

Figure~\ref{NMR} shows the Cu-NMR spectra of PrCu$_2$ powder at
selected temperatures.
The 300~K spectrum is characterized by two main features centered
at 80 and 86~MHz, which are ascribed to the NMR central-line of
the $^{63}$Cu and $^{65}$Cu isotopes, respectively. Both isotopes
possess a 3/2 nuclear spin but the gyromagnetic ratio of $^{65}$Cu
is roughly 7\% larger than that of $^{63}$Cu. We observe no sign
of any contribution from the $^{141}$Pr nuclei (reference
frequency $\sim 92$~MHz at 7.049~T), most probably because their
extremely large hyperfine coupling\cite{AFmK2} results in very
short relaxation times. Both Cu NMR lines are split by $\sim
1$~MHz. We ascribe this splitting to a second-order quadrupolar
perturbation of the Zeeman Hamiltonian. The wings observed between
75 and 90~MHz correspond to the first-order quadrupolar
perturbation. These wings are more prominent at lower temperatures
and it may be noted that, rather than a sharp edge, they exhibit a
smooth edge suggesting, as explained below, a non-zero anisotropy
$\eta$ of the electric field gradient (EFG). With decreasing
temperatures, the complexity of the spectrum increases. As may be
seen in fig.~\ref{NMR}, the asymmetry and width of the lines grows
and they partially merge together. The quadrupolar wings are
strongly affected as well and a large overall shift of the lines
towards higher frequencies is observed, indicating a progressively
increasing Knight-shift (K). The strongly asymmetric line-shape
suggests a large K-anisotropy.

The very complex structure of the low temperature NMR spectra,
caused by the simultaneous EFG and K anisotropies, makes a
quantitative analysis of the spectra very difficult. Therefore, we
keep the interpretation mostly at a qualitative level. To this end
we performed several simulations of the $^{63,65}$Cu NMR
powder-spectrum for different values of the microscopic parameters
characterizing K and EFG. The simulations involve a full
diagonalization of the nuclear Hamiltonian comprising the sum of
the quadrupolar and the Zeeman term corrected by K. The spectrum
is then obtained from the calculated transition energies and
intensities, averaged over random crystal orientations in order to
simulate a powder spectrum. The microscopic parameters
characterizing the quadrupolar Hamiltonian are the largest
component $V_{zz}$ of the EFG and the anisotropy $\eta$ in the
$xy$ plane. The parameters describing K are the isotropic and
anisotropic components $K_{\textrm{iso}}$ and $K_{\textrm{ani}}$,
respectively, as well as an axial-asymmetry term $\epsilon$. The
notation is standard and exemplified, for instance, in
Ref.~\onlinecite{carter} (pages 62-64).
\begin{figure}[!tb]
\center
\includegraphics[width=8cm]{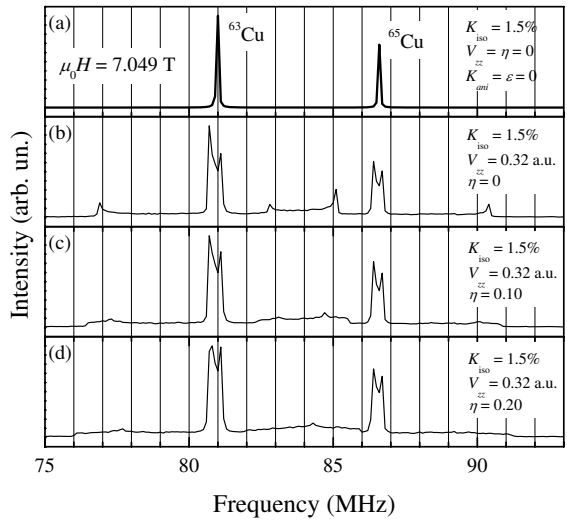}
\caption{Simulated $^{63,65}$Cu-NMR powder spectra in a field of
7.049~T with $K_\textrm{iso}=1.5\%$ and for different values of
the EFG parameters (see text).} \label{sim1}
\end{figure}
Figure~\ref{sim1}(a) shows the simulated spectra that we obtained
by considering only an isotropic $K_{\textrm{iso}}=1.5\%$ and no
quadrupolar contribution. Naturally, this spectrum consists of
only one line for each isotope, slightly shifted with respect to
the reference frequency. Figure~\ref{sim1}(b) was obtained by
adding an EFG with $V_{zz}=0.32$~a.u. and $\eta=0$. It is clear
that this term induces the appearance of the wing structures and a
splitting of the central lines. The effect of adding a non-zero
anisotropy $\eta$ is demonstrated in figs.~\ref{sim1}(c) and (d)
where the calculations with $\eta=0.10$ and $\eta=0.20$,
respectively, are displayed. Upon enhancing $\eta$, the sharp edge
of the wings is progressively smeared out, leading to a simulated
spectrum with a shape resembling the experimental data at 300~K.
The overall comparison of the experimental data with the
simulations of fig.~\ref{sim1}(b-d) suggests that the actual
components of the EFG acting at the copper site in PrCu$_2$ are of
the order of $V_{zz}=0.32$~a.u. and $\eta=0.10$, at least down to
100~K. An \textit{ab initio} calculation in which the internal
coordinates of the 12 ions in the unit cell of PrCu$_2$ were
optimized assuming the experimental lattice parameters
($a=4.4087$~\AA, $b=7.0551$~\AA, and $c=7.4441$~\AA)\cite{Jorge}
using the DMol$^3$ code,\cite{abini} yielded $V_{zz}=0.48$~a.u.
and $\eta =0.11$. Although the calculated absolute value of the
EFG is 50\% larger than that obtained by comparing experiment and
simulations as described above, the direction of the dominant
component of the EFG tensor is clearly established to lie in the
$bc$ plane. This is important for the discussion in
section~\ref{relax}.

\begin{figure}[!tb]
\center
\includegraphics[width=8cm]{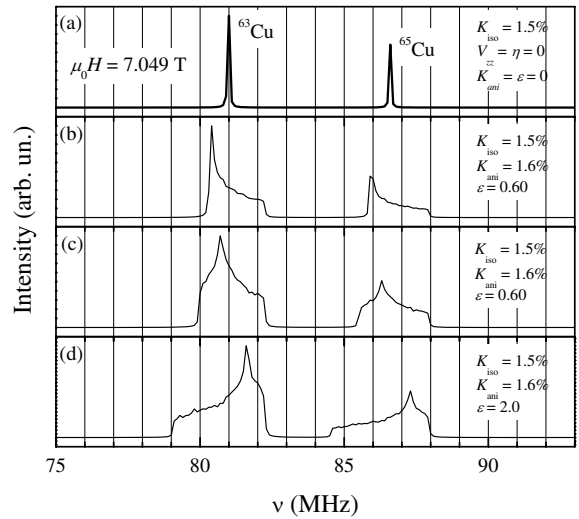}
\caption{Simulated powder $^{63,65}$Cu-NMR spectra at a field of
7.049~T and for different values of the K parameters (see text).}
\label{sim2}
\end{figure}

Figure~\ref{sim2}(b) shows a simulated spectrum where the
isotropic shift of $1.5\%$ is augmented by an anisotropic
component $K_{\textrm{ani}}=1.6\%$ but no axial-asymmetry. The
non-zero $K_{\textrm{ani}}$ induces an asymmetric broadening of
the central lines keeping rather sharp edges. The $\epsilon$ value
is responsible for the position of the maximum with respect to the
edges, as shown in figs.~\ref{sim2}(c) and (d), where the
simulated spectra with $\epsilon=0.6$ and $\epsilon=2.0$ are
shown, respectively.

Although these simulations allow for a qualitative description of
the main features of the NMR spectra, the situation turns out to
be much more complicated when we try to simultaneously take into
account the influence of both EFG and K. These two contributions
are indeed both non-negligible and, in principle, temperature
dependent. Moreover the corresponding tensors are usually non
diagonal in the crystalline-axes frame and their main frames, i.e.
the frames in which the two tensors are diagonal, are not
necessarily oriented along the same directions. The local symmetry
of the Cu-sites provides one constraint for the EFG and K main
frames. They should have one principal axis parallel to $a$.
Concerning the orientation of the other two axes in the $bc$
plane, nothing can be concluded from considering the
crystal-structure. Therefore the angle between the K and EFG main
frames in the plane is unknown. All the above considerations lead
to the conclusion that the influences of both K and EFG on the
spectral features cannot be regarded independently.

In spite of the difficulties described above, it is still possible
to extract some quantitative information from the NMR spectra
shown in fig.~\ref{NMR}. First of all, we can estimate the
quadrupolar frequency $\nu_Q$ of $^{63}$Cu from the widths of the
wings. The data imply $\nu_Q=8-9$~MHz with no significant change
upon decreasing temperature, at least down to 50~K. Inserting
$V_{zz}=0.32$~a.u. and $\eta=0.10$ into the relation
\begin{equation}
\nu_Q = \frac{3eV_{zz}Q}{2hI(2I-1)}\sqrt{1+\eta^2/3}, \label{nuQ}
\end{equation}
where $Q$ is the nuclear quadrupole moment and $I$ is the nuclear
spin, we obtain $\nu_Q = 8.3 \pm 0.8$~MHz. The second quantity
that we can estimate is the largest principal component
$K_\textrm{max}$ of the K tensor. In the presence of a K
anisotropy, the spectrum adopts the typical shape displayed in
figs.~\ref{sim2}(b)-(d).\cite{carter} The edges of the asymmetric
broadening occur at frequencies $(1 + K_\textrm{min})\nu_0$ and
$(1 + K_\textrm{max})\nu_0$, where $K_\textrm{min}$ is the
smallest component of the K tensor and $\nu_0$ is the reference
frequency of the considered nucleus.\cite{carter} The position of
the resonance maximum is at $(1 + K_\textrm{int})\nu_0$, where
$K_\textrm{int}$ is the third (intermediate) K-tensor
main-component.\cite{carter} The temperature dependence of the
high-frequency edge can be followed down to $\sim 20$~K, thus
allowing us to estimate $K_\textrm{max}(T)$.

We assume that the hyperfine field
$\roarrow{B}_\textrm{Cu}^\textrm{hf}$ acting at the Cu-site is
mainly induced by the polarization of the Pr magnetic-moments
$\roarrow{\mu}_\textrm{Pr}$ via an interaction mediated by the
conduction electrons, such that
\begin{equation}
\roarrow{B}_\textrm{Cu}^\textrm{hf}=\tensor{A}\roarrow{\mu}_\textrm{Pr},
\label{hyper}
\end{equation}
where $\tensor{A}$ is the tensor associated with the transferred
hyperfine field. The orientation of one principal axis of the K
tensor along the $a$ axis and the fact that it represents the easy
magnetization direction\cite{chi} indicate that the maximum K
component most likely lies along $a$, as was also suggested by our
preliminary measurements on the single crystal (data not shown
here). The same is implied by our estimate of $K_\textrm{max}(T)$,
which is found to be proportional to the susceptibility
$\chi_a(T)$ along the $a$ axis. For a given external magnetic
field $\roarrow{B}_0$, $\roarrow{B}_\textrm{Cu}^\textrm{hf}$ and
$\roarrow{\mu}_\textrm{Pr}$ can directly be obtained from K and
the susceptibility, respectively. Using eq.~\ref{hyper} we can
calculate the diagonal term of $\tensor{A}$ along $a$, namely
\begin{equation}
A_{aa}=K_a(T)/\chi_a(T),
\end{equation}
which is, of course, temperature independent and has a value of
$0.54 \pm 0.02$~T$/\mu_\textrm{B}$. This value is very useful in
the discussion of the NQR data.

A temperature-independent $A_{aa}$ and the fact that we do not
observe significant changes of the NMR spectrum around 65~K are
not compatible with an ordering of the Pr-moments at that
temperature. Because of the interaction between the Pr-moments and
the conduction electrons, any alignment of the Pr moments is
expected to have a strong influence on the Cu NMR response.

\section{NQR measurements}
\subsection{Spectra}\label{spectra}
\begin{figure}[!tb]
\center
\includegraphics[width=8cm]{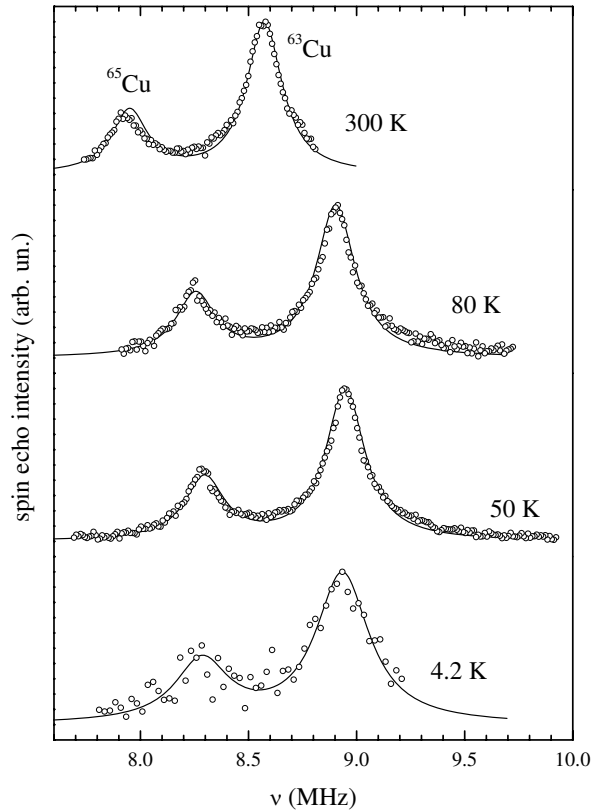}
\caption{$^{63,65}$Cu-NQR spectra of PrCu$_2$ at selected
temperatures (symbols). Solid lines are best-fit curves to the
data (see text).} \label{NQR}
\end{figure}

\begin{figure}[!tb]
\center
\includegraphics[width=8cm]{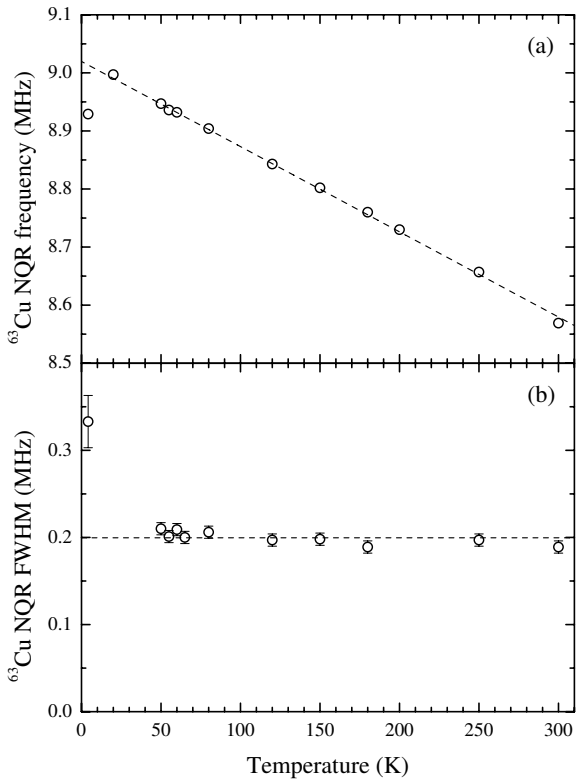}
\caption{Symbols: (a)$^{63}$Cu-NQR frequency and (b) line-width of
PrCu$_2$ as a function of temperature, as obtained from the
best-fits to the NQR spectra. The dashed lines represent (a) a
linear fit and (b) the average of the data above 7~K.} \label{fw}
\end{figure}

The rather large value of $\nu_Q$ indicated by the NMR data
suggests that an NQR signal can be observed in an accessible
frequency range. Figure~\ref{NQR} shows Cu NQR spectra at selected
temperatures.
The recorded data reflect the expected simple NQR spectrum with
one single line per isotope, reflecting the respective $\pm 1/2
\rightarrow \pm 3/2$ transitions. The relative peak positions of
the $^{63}$Cu and $^{65}$Cu signals are consistent with the
nuclear quadrupole moment, which is roughly 8\% larger for
$^{63}$Cu than for $^{65}$Cu. From the raw data it is clear that
the $^{63}$Cu NQR frequency is within 10\% of the estimate of
$\nu_Q$ obtained from the NMR spectra. The NQR frequency increases
with decreasing temperature, whereas no major changes are observed
in the $T$-dependence of the line-widths. In order to obtain more
quantitative information, we applied a best-fit procedure that is
based on the frequency dependence of the intensity $S(\nu)$ given
by
\begin{eqnarray}
\nonumber
S(\nu)& = & \frac{A\alpha_{65}}{(\nu-\nu_Q / \beta)^2+(\Gamma /2)^2}+\\
& & \frac{A\alpha_{63}}{(\nu-\nu_Q)^2+(\Gamma /2)^2},
\label{fitNQR}
\end{eqnarray}
where $A$ is the overall intensity, $\alpha_{65,63}$ are the
natural abundances of $^{65,63}$Cu, $\nu_Q$ is the $^{63}$Cu NQR
frequency, $\Gamma$ is the line-width (assumed to be equal for
both isotopes), and $\beta=1.078$ is the ratio between the
quadrupolar moments of $^{63}$Cu and $^{65}$Cu. The experimental
intensity $I(\nu,T)$ is
\begin{equation}
I(\nu,T)=S(\nu)\nu \left[1-\exp\left(-\frac{h\nu}{k_B
T}\right)\right],
\end{equation}
where the multiplicative factor $\nu$ is due to the fact that our
measurement is inductive; the term in square brackets represents
the Boltzmann population factor. With this procedure, we fitted
all the spectra using only $A$, $\nu_Q$, and $\Gamma$ as free
parameters. Physically relevant are the latter two and we plot
them as a function of temperature in fig.~\ref{fw}.

The frequency $\nu_Q$ increases linearly with decreasing
temperature down to $T_\textrm{JT}$, where a sudden drop in
$\nu_Q(T)$ is observed. This reflects the JT structural transition
which induces a sudden change in the EFG and thus in $\nu_Q$. The
overall change of $\nu_Q$ between 300~K and 20~K is about 5\%.
Since $\nu_Q$ is related to both $V_{zz}$ and $\eta$ by means of
eq.~\ref{nuQ}, it is not possible to sort out whether the observed
change in $\nu_Q$ is to be ascribed to $V_{zz}$, or $\eta$, or
both. Nevertheless, if we assume a constant $V_{zz}$, the
$\nu_Q(T)$ values imply that the anisotropy should change from
$\eta \sim 0.10$ at 300~K to approximately 0.60 at 20~K. Such a
dramatic change in the asymmetry is very unlikely because it would
require a strong modification of the electronic density, which
does not seem to be compatible with the small anisotropy of the
thermal expansion coefficients.\cite{ThermExp} Moreover no
evidence of such a large variation can be inferred from the NMR
data shown in fig.~\ref{NMR}, because in the presence of a large
$\eta$ the edges of the quadrupolar wings are dramatically smeared
out. It is therefore reasonable to ascribe most of the change in
$\nu_Q$ to a corresponding increase of $V_{zz}$, probably induced
by the thermal contraction of the lattice.\cite{ThermExp} The lack
of any anomaly in $\nu_Q(T)$ around 65~K does not add support to
the magnetic-order scenario inferred from the $\mu$SR data.

Between room-temperature and $T_\textrm{JT}$ the line-width
$\Gamma$ is almost constant. In particular, also $\Gamma (T)$
exhibits no anomaly around 65~K. Likewise the absolute values of
$\Gamma$ are in contrast with a magnetic order of the Pr moments.
Because of the interaction of the Pr-moments with the conduction
electrons, magnetic order among them would imply a local field
$\roarrow{B}_\textrm{Cu}^\textrm{hf}$ acting on the Cu-sites. As
discussed above (sect.~\ref{NMRmeas}), the contribution to this
field of any Pr moment oriented along $a$ can be calculated by
means of the corresponding hyperfine constant, i.e.
$B_\textrm{Cu}^\textrm{hf}=A_{aa} \mu_\textrm{Pr}$. We calculated
the local magnetic fields acting on Cu sites by invoking the six
nearest-neighbors Pr moments, adopting the claimed AF
configuration with an incommensurate modulation of the moments.
The corresponding vector $\roarrow{q}=(0.24, 0, 0.68)$ and the
modulation amplitude is 0.29~$\mu_\textrm{B}$.\cite{muons} We
obtain a distribution of $B_\textrm{Cu}^\textrm{hf}$ ranging from
0 to 0.06~T. Such a distribution of fields should be reflected in
a broadening $\Delta \nu = 1.4$~MHz of the NQR line. This value is
almost one order of magnitude larger than $\Gamma$ of our signals,
thus indicating that a magnetic order of the claimed
type\cite{muons} is incompatible with our data.

Since the features extracted from the NQR spectra do not confirm,
but rather rule out AF ordering of the Pr moments at 65~K, it
seems in order to discuss possible causes for this discrepancy. At
present we consider two scenarios which we discuss in more detail
in section~\ref{muvsNMR}. The first is based on the assumption
that the apparent order observed with muons is mainly due to a
muon-induced local enhancement of slow spin fluctuations. As a
second possibility we consider that the observed order is not
static, but fluctuates on a time-scale of the order of 1~$\mu$s.
With $\mu$SR and the related precession frequency of the order of
10~MHz,\cite{muons} a non-zero field would be indicated. The
characteristic time period in employing the NMR/NQR spin-echo
technique are in the range of tens of microseconds and therefore
the field fluctuations in the MHz range would be averaged to zero
and thus would not affect the resonance signal. Under these
circumstances, $\mu$SR would indicate a ``static'' magnetic
correlation with a distribution of internal fields, whereas in
NMR/NQR the fields would be averaged to zero. Such a slow dynamics
has no simple explanation. Of course, the presence of a temporally
fluctuating field should be reflected in the dynamical magnetic
properties on a microscopic scale, namely the SLRR $T_1^{-1}$ and
the SSRR $T_2^{-1}$. At any rate, the onset of magnetic-order is,
usually in the vicinity of the transition, accompanied by strong
magnetic fluctuations, which should be reflected in the relaxation
rates.

\subsection{Relaxation} \label{relax}
\begin{figure}[!tb]
\center
\includegraphics[width=8cm]{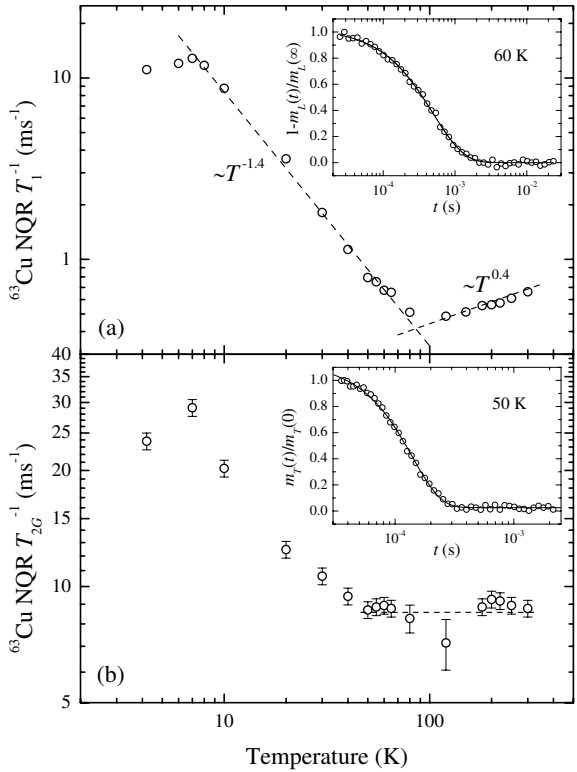}
\caption{(a) Spin-lattice and (b) spin-spin relaxation rates as a
function of temperature. Dashed lines in (a) are power-law fits to
the data in the temperature range above and below 100~K. The
uncertainty of the data points in (a) is of the order of the
diameter of the open circles. The horizontal dashed line in (b) is
a guide to the eyes. Insets: examples of the time dependent
longitudinal (a) and transversal (b) magnetization. Solid lines in
the insets of (a) and (b) are best fits to data according to
equations \ref{T1} and \ref{T2}, respectively.} \label{T1T2}
\end{figure}

With the procedures described in section~\ref{expt}, we measured
the time recovery of the longitudinal $m_L(t)$ and transversal
$m_T(t)$ magnetization. Representative examples of $m_L(t)$ and
$m_T(t)$ are shown in the insets of fig.~\ref{T1T2}(a) and (b),
respectively.
All the $m_L(t)$ curves can be fitted by means of the standard
exponential function that applies for $I=3/2$,\cite{fitT1} namely
\begin{equation}
m_L(t)=m_L(\infty) \left[ 1-\exp\left( -\frac{\rho t}{T_1}\right)
\right], \label{T1}
\end{equation}
where $m_L(\infty)$ is the equilibrium magnetization, $T_1$ is the
spin-lattice relaxation time, and $\rho$ is an $\eta$-dependent
parameter, which can be approximated with $\rho \simeq 3$ for
$0\lesssim \eta \lesssim 0.2$.\cite{fitT1} Because the standard
dipole-dipole interaction is not the main spin-spin relaxation
mechanism, $m_T(t)$ cannot be fitted by means of a simple
exponential function. The stronger indirect interaction mediated
by the conduction electrons via the Ruderman-Kittel-Kasuya-Yosida
mechanism leads to a gaussian correction\cite{T2RKKY,GaussT2} such
that
\begin{equation}
m_T(t)=m_T(0) \exp \left[-\frac{t}{T_R}-\frac12 \left(
\frac{t}{T_{2G}} \right)^2 \right], \label{T2}
\end{equation}
where $m_T(0)$ is the initial transverse magnetization, $T_{2G}$
is the gaussian spin-spin relaxation time, and $T_R=3T_1/(2+r)$ is
a time-scale depending on the SLRR and on its anisotropy $r$.
Since this anisotropy is expected to be not as large as, e.g., in
the layered copper oxides ($r \sim 4$), we assume $r \sim 1$ and
hence $T_R \sim T_1$. As long as $T_R$ is significantly larger
than $T_{2G}$, their values are quite independent. Our assumption
$T_R = T_1$ is thus justified \textit{a posteriori}, both by the
good quality of the fits and by the fact that we obtain
systematically $T_{2G}<T_1$.

From the fits based on equations \ref{T1} and \ref{T2}, we obtain
the temperature dependencies of the SLRR and SSRR, as shown in
figs.~\ref{T1T2}(a) and (b), respectively. Neither $T_1^{-1}(T)$
nor $T_{2G}^{-1}(T)$ exhibit the expected features that would
reflect an onset of magnetic-order around 65~K. The fluctuations
that accompany such a transition should cause a distinct
enhancement of the relaxation rates, resulting in a peak in
$T_1^{-1}(T)$ around the ordering temperature. From
fig.~\ref{T1T2}(b) we take it that $T_{2G}^{-1}(T)$ is constant
down to around 40~K where it starts to increase with decreasing
temperature, displaying a peak at $T_\textrm{JT}$. The decrease of
$T_{2G}^{-1}(T)$ below $T_\textrm{JT}$ most likely reflects the
reduction of the quadrupolar fluctuations. Also $T_1^{-1}(T)$
peaks at $T_\textrm{JT}$, but the behavior at higher temperature
is more difficult to interpret. The temperature dependence of the
SLRR is neither compatible with a dominating relaxation via
conduction electrons ($T_1^{-1}(T) \sim T$), nor a relaxation via
local magnetic moments ($T_1^{-1}=\textrm{const.}$) or by
spin-wave type excitations ($T_1^{-1} \sim T^3$). Instead
$T_1^{-1}(T)$ exhibits a power-law type behavior characterized by
a negative exponent (-1.4) between 8 and 100~K and a positive one
(0.4) above 100~K. In order to interpret this puzzling behavior,
we tried to determine the leading spin-lattice relaxation
mechanism by measuring $T_1$ also for $^{65}$Cu. With
$^{63,65}T_1$, $^{63,65}\gamma$, and $^{63,65}Q$ we denote the
spin-lattice relaxation time, the gyromagnetic ratio, and the
quadrupole moment of $^{63,65}$Cu, respectively. If the dominant
mechanism is magnetic,
\begin{equation}
^{63}T_1/^{65}T_1=(^{65}\gamma /^{63}\gamma)^2=1.167,
\end{equation}
whereas for a relaxation driven by the quadrupolar interaction we
expect
\begin{equation}
^{63}T_1/^{65}T_1=(^{65}Q /^{63}Q)^2=0.872.
\end{equation}
We established that below 100~K $^{63}T_1/^{65}T_1 \simeq 1.1$
which indicates that the relaxation is predominantly of magnetic
origin. At higher temperature we find $^{63}T_1/^{65}T_1 \simeq
1$, suggesting that the magnetic and quadrupolar interactions are
of comparable importance. The fact that the magnetic fluctuations
dominate below 100~K again suggests that the internal field, if
any is present, is not static. It is also worth mentioning that
the puzzling $T_1^{-1}(T)$ cannot be ascribed to the temperature
dependence of the population of the Pr levels in simple terms.
Such a case would require that the lifetimes of these levels are
different for each of them. Bearing in mind the single-exponential
behavior exhibited by all our $m_L(t)$ curves, this possibility
appears as rather unlikely.

\section{Comparison between NMR/NQR and $\mu$SR
measurements}\label{muvsNMR} Our \textit{ab initio} calculations
indicate that the dominant axis of the EFG and thus the
quantization direction in NQR, is orthogonal to the $a$ axis (see
section~\ref{NMRmeas}). If we assume that the internal field along
$a$ observed by $\mu$SR is not static but fluctuates, this would
result in a large SLRR whereas it should have only minor effects
on the SSRR. This is consistent with our observation that on
decreasing the temperature, $T_1^{-1}(T)$ increases much more
rapidly than $T_{2G}^{-1}(T)$. As already mentioned
(sect.~\ref{spectra}), this scenario could reconcile the NMR/NQR
data with the $\mu$SR results, provided that the time-scale of the
spin polarization is long ($\sim 1$~$\mu$s). This would also be
consistent with the observation of a gaussian transverse
relaxation, since the presence of such a term requires an
enhancement of the low-frequency components of the local
susceptibility,\cite{GaussT2} i.e., a slowing down of the spin
dynamics.

\begin{figure}[!tb]
\center
\includegraphics[width=7cm]{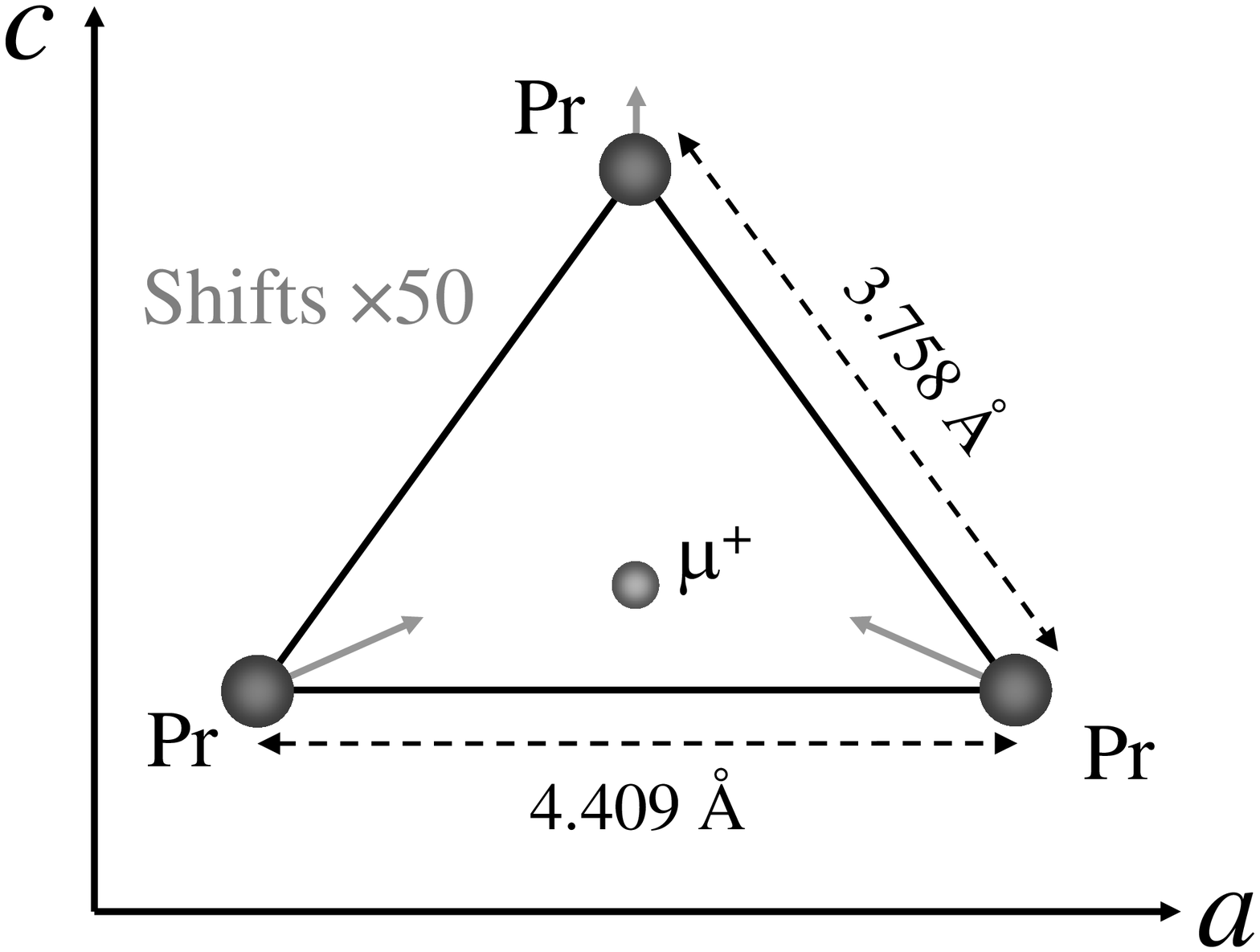}
\caption{Sketch of the positions of the Pr ions surrounding the
muon, as obtained from the \textit{ab initio} calculations (see
text). The arrows indicate the displacement of the Pr ions
(multiplied by 50) induced by the presence of the muon.}
\label{muonPr}
\end{figure}

In order to reconcile the results of the NMR/NQR and $\mu$SR
experiments, we also consider the possibility that the magnetic
order observed by $\mu$SR is mimicked by a quasi-static local
enhancement of moment polarization by the probe itself. The muon,
considered as a positive point charge, will attract electrons and
thus lead to a rearrangement of charge and possibly of the ion
positions. If, because of a slow enough spin fluctuation, the
intrinsic electronic charge distribution is spin polarized during
the duration of the muon lifetime, the muon will experience a spin
density exceeding the one induced by the fluctuation. For our
calculations\cite{MuonCal} we introduce a hydrogen
atom\cite{Hydrog} at the position of the muon derived by Schenck
et al. $(1/2,1/4,0.6406)$\cite{muons} in every second cell along
the $a$ direction, which amounts to one muon per 8 formula units,
and let all atoms relax at fixed volume. The resulting coordinates
for the position of the muon are $(1/2,1/4,0.6465)$, independent
of the imposed spin polarization between 0 and 1~$\mu_\textrm{B}$
per unit cell. The Pr-ions in the nearest neighbor triangle are
displaced as shown in fig.~\ref{muonPr}.
The Pr-Pr distance is reduced by $\sim 0.3$\% which corresponds to
a compression by approximately 5~kbar.\cite{AFHP1} At this point,
it is worth mentioning that at external pressures exceeding
12~kbar, AF ordering of the Pr moments has been observed for
$T\lesssim 9$~K,\cite{AFHP1,AFHP2} thus indicating that a lattice
compression enhances the magnetic correlations. Imposing a spin
polarization of 1~$\mu_\textrm{B}$ per unit cell, which
corresponds to an average spin density of $6.4 \cdot
10^{-4}$~a.u., leads to a spin density at the muon site which is a
factor 6.6 larger than the average one. Reducing the imposed spin
polarization by a factor 4 yields a spin density at the muon 12
times larger than the average value. One might argue that the
conclusions drawn from a calculation in which a hydrogen atom is
placed in every second cell cannot be applied to the analysis of
$\mu$SR experiments, which are performed with isolated muons. Our
experience shows that if we double the number of hydrogens to one
per cell, the enhancement of the spin density at the muon-site is
reduced by 15\%. In other words, a larger dilution is expected to
enhance the influence of the muon. So there is no doubt that the
presence of the muon affects the response of the system, namely a
small spin polarization in the system will be amplified at the
muon site. A quantitative assessment of the temperature dependence
of this extrinsic effect is out of reach of the presently used
calculation method.

\section{Conclusions}
The results of our NMR/NQR study of both the static (spectra) and
dynamical (relaxation rates) microscopic magnetic properties of
PrCu$_2$ provide no evidence for the onset of any magnetic order
around 65~K that was previously claimed from $\mu$SR
data.\cite{muons} As argued above, the $\mu$SR results may not
reflect magnetic order but are rather due to a muon-induced local
enhancement of slow polarization fluctuations. If the fluctuation
has a time scale of the order of 1~$\mu$s, it could, as argued in
section~\ref{spectra}, appear as static in $\mu$SR and vanishing
in NMR/NQR. The origin of the anomalous power-law behavior of the
SLRR is not yet understood, but it clearly indicates that the
magnetic behavior of PrCu$_2$ is far from that of a simple Pauli,
Curie-Weiss type, or spin-wave system.

The present results call for further studies in order to clarify
the complex magnetic properties of PrCu$_2$. In particular it will
be important to carry out NMR/NQR measurements in an AF phase.
This could be obtained either by means of measurements below 54~mK
or at pressures exceeding 12~kbar, where AF ordering among
localized Pr moments has been observed for $T\lesssim
9$~K.\cite{AFHP1,AFHP2}

\begin{acknowledgments}
We thank A. Schenck for a detailed discussion of this work and we
wish to acknowledge valuable contributions from J. Hinderer, S.
Weyeneth, P. W\"agli, E. Fischer, R. Helfenberger, H. R.
Aeschbach, and F. N. Gygax. This research benefitted from
financial support of the NCCR program MaNEP of the Schweizerische
Nationalfonds f\"ur wissenschaftliche Forschung.
\end{acknowledgments}


\begin{thebibliography}{99}
\bibitem{MuonCal} In these calculations, the Pr $4f$ electrons were
considered as part of the (rigid) ionic core, and the Pr valence
electrons were described by a pseudopotential, since we were not
interested in the density or spin-density in the immediate
neighborhood of the Pr nucleus.
\bibitem{Hydrog} As far as static properties are concerned, a
positive muon or a proton are equivalent. The choice of a neutral
atom is justified by the efficient screening of the point charge
by the surrounding electrons and avoids the introduction of an
artificial neutralizing background in the total energy
calculation. Nevertheless, we also performed a calculation in
which the hydrogen atom is replaced by a positive point charge.
The equilibrium position remains the same within 0.001 Å and the
enhancement of the spin density is increased by 3 percent. Hence
our order of magnitude estimate of the effect should be valid,
irrespective of the charge state of the introduced object
representing the muon.
\end{thebibliography}
\end{document}